\RequirePackage{fix-cm}
\documentclass[conference,letterpaper]{IEEEtran}
\usepackage[T1]{fontenc}
\usepackage{cite}
\usepackage{amsmath,amssymb}
\usepackage{graphicx}
\usepackage{float}
\usepackage{booktabs,array,multirow}
\divide\textfloatsep by 4
\divide\dbltextfloatsep by 4
\divide\intextsep by 3

\graphicspath{{figures/}}
\newcommand{\sys}{QuarkDB}
\newcommand{\tablesize}{\fontsize{8}{9.6}\selectfont}
\usepackage{listings}
\usepackage[hidelinks]{hyperref}
\hypersetup{
  pdftitle={QuarkDB: A Hybrid Database System for Quantum Data Management},
  pdfauthor={Wenrui Zhang, Yu Wang, Xiao Zhang},
  pdfkeywords={quantum data management, hybrid query engine, execution engine, storage engine, QuarkDB}
}
\lstdefinestyle{quansical}{
  basicstyle=\ttfamily\footnotesize,
  columns=fullflexible,
  keepspaces=true,
  breaklines=true,
  breakatwhitespace=false,
  showstringspaces=false,
  linewidth=\linewidth,
  frame=single,
  framerule=0.4pt,
  framesep=3pt,
  xleftmargin=4pt,
  xrightmargin=4pt,
  aboveskip=4pt,
  belowskip=4pt
}

\begin{document}
\bstctlcite{IEEEcontrol}
\title{\sys{}: A Hybrid Database System for Quantum Data Management}
\author{
\IEEEauthorblockN{Wenrui Zhang}
\IEEEauthorblockA{Renmin University of China\\wrzhang@ruc.edu.cn}
\and
\IEEEauthorblockN{Yu Wang}
\IEEEauthorblockA{Renmin University of China\\wangyuuu1@ruc.edu.cn}
\and
\IEEEauthorblockN{Xiao Zhang}
\IEEEauthorblockA{Renmin University of China\\zhangxiao@ruc.edu.cn}
}

\maketitle
\begin{abstract}
Emerging quantum applications like quantum communication generate large volumes of high-dimensional quantum data, requiring efficient data analysis and storage. However, classical database systems often struggle to meet requirements such as quantum processing or version management. This paper presents \sys{} designed for efficient quantum data management. \sys{} establishes a unified object model for relational and quantum data and defines the syntax of Quantum Structured Query Language (QSQL). To support efficient quantum data management, \sys{} integrates three key components: (1) a Hybrid Query Engine with cost-based hybrid plan selection over classical and quantum operators, (2) an Execution Engine with a three-level instructions intermediate representation, QuarkDB-IR, to support quantum execution and reduce quantum processing costs, and (3) a Storage Engine supporting quantum data version management, state reconstruction and data reuse. Compared with classical databases such as SQLite and PostgreSQL or quantum databases like Qute, our \sys{} achieves query-latency speedups ranging from $2.14\times$ to $25.86\times$ under extensive real-world workloads.
\end{abstract}
\section{Introduction}
\label{sec:introduction}
Quantum computing has been applied to several research fields, including communication networks~\cite{netsquid}, cryptographic protocols~\cite{quantumcryptography}, and combinatorial optimization~\cite{qaoa}, and offers performance advantages for specific computational tasks~\cite{trummer2016,grover,huang2022}. However, these applications also generate substantial quantum data, including circuit descriptions, measurement outcomes, run records and so on~\cite{hai2025}.  

\begin{figure}[!htbp]
 \centering
 \includegraphics[width=\columnwidth]{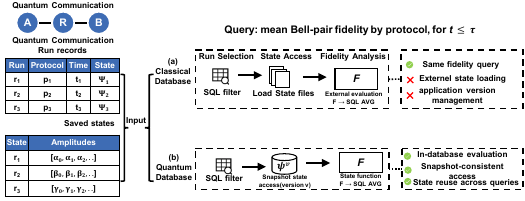}
 \caption{Entanglement-Fidelity Queries in Classical and Quantum Databases}
 \label{fig:simulation-example}
\end{figure}

Compared to classical data, quantum data carries complex amplitudes and correlations across qubits. Storing quantum states as state vectors requires storage that grows exponentially with the number of qubits~\cite{hai2025}, posing complex challenges for relational databases. Meanwhile, it is also particularly difficult to analyze quantum data in a classical database. For example, Figure~\ref{fig:simulation-example} illustrates the entanglement distribution through a repeater chain in quantum communication, where a common analysis is computing mean Bell-pair fidelity by protocol for runs completing within a time limit~\cite{netsquid}. If the application uses the classical database to process it, it must load state files into a simulator and coordinate state versions with query results. Repeated loading adds the cost of encoding and decoding, while independently storing complete versions increases storage. For example, a dense 30-qubit state requires 16~GiB to store~\cite{qulacs}, which means retaining its 100 versions data requires approximately 1.56~TiB, imposing substantial storage requirements. These requirements motivate the development of quantum databases to manage quantum data and meet the needs of future quantum applications. The lower part of Figure~\ref{fig:simulation-example} describes an ideal quantum database design, wich integrates state evaluation with relational query processing and provides snapshot-consistent access to states and their associated records. Managing state versions within the database supports reuse across queries, reducing repeated state loading and redundant storage.

Recent quantum database research often adopts the quantum augmented architecture of database system, using quantum solvers to accelerate database optimization~\cite{qdbo}. QDBO, DEDALUS, and Q$^2$O apply this design to join ordering and related optimization problems~\cite{qdbo,dedalus,q2o2025}. However, supporting quantum operations alongside relational queries introduces three additional challenges. (1) \emph{Complexity in selecting candidate plans.} Existing studies usually explore quantum algorithms to improve database performance,
with many focusing on optimization problems~\cite{nayak2026,fankhauser2023mqo,nayak2024split,liu2025bushy,calikyilmaz2023opportunities}. But the combination of quantum and classical operators additionally introduces the cost of selecting the proper operator for queries, multiplying the plan alternatives, so choosing operators independently can produce an expensive complete plan~\cite{qute}. (2) \emph{Coordinated hybrid execution.} Classical relational operators and quantum operations differ in data representations, instruction requirements, and result processing. Coordinating them within a database requires compatible execution interfaces, while repeated representation conversions and instruction compilation introduce additional overhead~\cite{hopf2026,burgholzer2026}. Meanwhile redundant quantum backend instruction processing and circuit operations may further increase execution costs. (3) \emph{State version management and reuse.} Quantum states depend on computational histories and the relational tuple versions used in their preparation, which conventional tuple versioning alone cannot capture~\cite{tqsim,qtask}. Storing every state version as a complete state vector introduces redundant storage, while independently reconstructing historical states repeats shared circuit computation.

To address these problems, we propose \sys{}, which can manage relational data and quantum states on classical computers. With Structured Query Language (SQL) and Quantum Structured Query Language (QSQL), \sys{} can process queries combining relational operations and quantum procedures, optimize the plan candidate space via dynamic programming and select execution strategies based on resource constraints. It can also manage quantum states remaining accessible with their corresponding relational records, allowing subsequent queries to reuse these states.

In summary, we make the following contributions.
\begin{itemize}
\item We propose a new database system \sys{} for quantum data management with a hybrid architecture. By designing a unified object model with five fields, our system enables the unified description of both quantum and classical data. Additionally, we design QSQL for processing the quantum data.
\item We design a Hybrid Query Engine that optimizes classical and quantum operators combination using dynamic programming. We propose a cost model based on candidate cardinality and output cardinality that estimates candidate preparation, quantum search, result readout, and backend switching. By reusing cardinality and incremental-cost estimates across subplans, the hybrid query engine prunes infeasible alternatives and generates the lowest cost executable plans.
\item We implement an Execution Engine with a three-level instructions intermediate representation, QuarkDB-IR. With these instructions, our system can efficiently compose database operations on various quantum circuits and translate optimized programs into executable operations for quantum backends. \sys{} also introduces two execution optimization strategies, Instruction Rewriting and Circuit Reduction, to reduce the cost of instruction processing and circuit compilation.

\item We design a storage engine with strategies for version management, state reconstruction and quantum data reuse. Our version management strategy enables historical state reconstruction through replay from shared checkpoints or inverse execution from retained successors. Meanwhile, we propose a cost-based quantum data reuse strategy to reduce the cost of state preparation by selecting between state reconstruction and observable propagation.
\end{itemize} 

The paper is organized as follows: Section~\ref{sec:architecture} outlines the system architecture. The hybrid query engine, three-level execution engine, and storage engine are detailed in Sections~\ref{sec:query}--\ref{sec:state}. We present the evaluation and related work in Sections~\ref{sec:evaluation}--\ref{sec:related}, followed by conclusions in Section \ref{sec:conclusion}.

\section{System Overview}
\label{sec:architecture}

We first define the unified object model in Section~\ref{sec:data model}, and then introduce the architecture of \sys{} to show how quantum processing is seamlessly integrated into a classical database in Section~\ref{sec:\sys{}_architecture}. Finally we introduce the native QSQL grammar for using \sys{} in Section~\ref{sec:qsql_grammer}.

\subsection{Data Model}
\label{sec:data model}
\sys{} describes classical and quantum data using five fields: an identifier, an object type, version metadata, a payload, and references. Relations retain classical database schemas, column types, and constraints; relation versions and quantum objects share this description.

The identifier is a universally unique identifier (UUID) naming one object version. The object type distinguishes relation versions, states, circuits, and classical results, and determines the payload format. Version metadata records the initial version's identifier, the current version number, the predecessor's identifier, and creation time. The initial identifier groups related versions; predecessor links describe their derivation and branching.

The payload holds the content specified by the object type: relation identifiers and row counts, state amplitudes or reconstruction records, circuit gates and parameters, or observation values and sample counts. References record other object identifiers and their roles, linking an object to its inputs, circuit, and producing execution.

Applying a circuit creates a state with a new UUID and the same object type and initial version identifier as its input. Version metadata identifies the input as its predecessor. The payload describes the resulting state, while references identify the circuit and execution. Run tuples associate state identifiers with application metadata, as in Figure~\ref{fig:simulation-example}.

For a quantum state prepared from relational data, its qindex maps basis positions to logical row identifiers and the source tuple versions used during state preparation. \sys{} multiversion concurrency control (MVCC) determines visible tuple and catalog versions. A quantum database snapshot retains references to selected quantum data versions and relational tuple versions, together with their required dependencies, to reduce the cost of quantum state reconstruction.

\subsection{\sys{} Architecture}
\label{sec:\sys{}_architecture}

\begin{figure}[!htbp]
 \centering
 \includegraphics[width=\columnwidth]{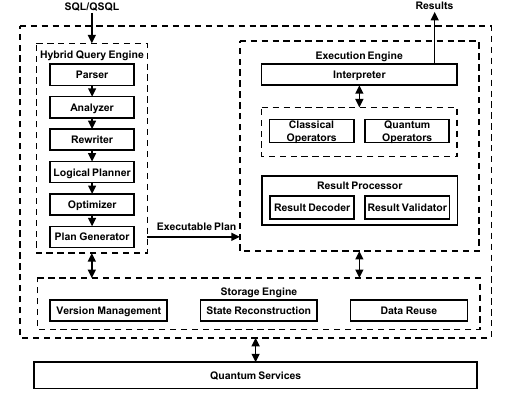}
 \caption{Architecture of \sys{}}
 \label{fig:architecture}
\end{figure}

\sys{} extends a classical relational database with a Hybrid Query Engine, an Execution Engine based on the three-level instruction intermediate representation (IR), and a Storage Engine for quantum data management. Figure~\ref{fig:architecture} shows their interaction. \sys{} uses a classical database for classical data management and transaction services, and a quantum processing backend for quantum circuit simulation~\cite{qulacs}.

\textbf{Hybrid Query Engine.} The parser accepts QSQL through sessions; the analyzer resolves relations and expressions against the catalog. The optimizer uses cardinality estimates and a cost model accounting for quantum preparation, search, readout, and cross-backend transitions. It compares fully classical and quantum-containing plans. A mixed plan can filter tuples before preparing the reduced input for quantum processing. Section~\ref{sec:query} describes query processing.

\textbf{Execution Engine.} Complete classical plans execute in the classical database backend. For plans containing Quantum operators, the compiler expands System Commands into Macro Instructions and then Execution Primitives. To reduce the cost of instruction execution and circuit compilation, \sys{} designs Instruction Rewriting to reduce repeated computation and instruction dispatch, while Circuit Reduction simplifies gate sequences. The interpreter invokes the quantum process backend for circuit execution. Section~\ref{sec:execution} describes the instruction hierarchy, compilation, and execution.

\textbf{Storage Engine.}
The storage engine manages quantum data versions and persistent database snapshots. Operation records track computational dependencies between quantum states, while qindexes associate states with their source tuple versions. Shared checkpoints support historical reconstruction through forward replay or inverse execution from retained successors. For historical queries, cost-based reuse selects between inverse reconstruction and observable propagation. Section~\ref{sec:state} describes these strategies.

\subsection{QSQL Grammar}
\label{sec:qsql_grammer}
We design QSQL to combine relational queries with operations on quantum states and their histories. QSQL extends SQL with commands and options for quantum processing while retaining relational expressions. Expressions can apply stored circuits, reconstruct states retained in snapshots, and evaluate observables on their amplitudes. Table~\ref{tab:qsql-grammar} summarizes QSQL statements. Brackets denote optional syntax, and braces list alternatives.

\begin{table}[!htbp]
\centering
\caption{QSQL Statements and Quantum Operations}
\label{tab:qsql-grammar}
\fontsize{8}{9.2}\selectfont
\setlength{\tabcolsep}{3pt}
\begin{tabular}{@{}p{0.20\columnwidth}
p{\dimexpr0.80\columnwidth-2\tabcolsep\relax}@{}}
\toprule
\textbf{Operation} & \textbf{Syntax} \\
\midrule

Query &
QSELECT [MEASURE] \textit{expressions}
[FROM \textit{relation expression}]
[WHERE \textit{predicate}]
[GROUP BY \textit{keys}]
[HAVING \textit{predicate}]
[ORDER BY \textit{order list}]
[LIMIT \textit{count}] [OFFSET \textit{count}]
[\textit{quantum clauses}] \\

Insert &
QINSERT INTO \textit{relation} [(\textit{columns})]
VALUES (\textit{values}) [, \ldots]
[RETURNING \textit{expressions}]
[\textit{quantum clauses}] \\

Update &
QUPDATE \textit{relation} [PERMUTE]
SET \textit{assignments}
[WHERE \textit{predicate}]
[RETURNING \textit{expressions}]
[\textit{quantum clauses}] \\

Delete &
QDELETE FROM \textit{relation}
[WHERE \textit{predicate}]
[RETURNING \textit{expressions}]
[\textit{quantum clauses}] \\
\midrule

Aggregation &
AGGREGATE MODE
\{QUANTUM\_COUNTING $|$ 
AMPLITUDE\_ESTIMATION $|$ DURR\_HOYER\}
[CONFIDENCE \textit{level}] \\

Readout &
SHOTS \textit{budget}
RETURN MODE
\{RELATION $|$ MEASUREMENT $|$ BOTH\} \\

Execution &
PLAN \{AUTO $|$ CLASSICAL $|$ QUANTUM\}
BACKEND \{QULACS $|$ AER $|$ WUKONG\} [EXACT] \\
\midrule

State expressions &
qstore.apply\_circuit(\textit{execution}, \textit{state}, \textit{circuit})\newline
qstore.snapshot\_restore\_state(\textit{snapshot}, \textit{state})\newline
qstore.state\_probability(\textit{state}, \textit{mask}, \textit{outcome})\newline
qstore.snapshot\_expectation(\textit{snapshot}, \textit{state}, \textit{observable}) \\
\bottomrule
\end{tabular}
\end{table}

As summarized in Table~\ref{tab:qsql-grammar}, QSQL retains the classical database query and modification clauses, adding PERMUTE to express updates as permutations of encoded values. AGGREGATE MODE defines the quantum procedure for aggregate evaluation, while CONFIDENCE defines the confidence level of aggregate intervals. SHOTS defines the sampling budget used by quantum operators to produce observations and estimates. RETURN MODE defines whether query output contains tuples, measurement records, or both; QSELECT MEASURE requests both. PLAN defines whether operator selection is cost-based, restricted to classical processing, or required to include a Quantum operator. BACKEND selects the circuit simulator, and EXACT requires complete results and exact aggregation. State expressions access quantum objects by identifier. apply\_circuit creates successor state versions, while snapshot\_restore\_state reconstructs saved amplitudes for historical reads. state\_probability returns the probability of a specified qubit outcome, and snapshot\_expectation returns an observable's expectation value, sampling variance, and sample count.

The following QSQL expression describes the Snapshot state access steps in Figure~\ref{fig:simulation-example}. It accesses the saved quantum state of a selected run and evaluates bell\_projector, the target Bell-state projector for endpoints A and B. The resulting fidelity $F$ provides this run's contribution to the subsequent SQL AVG computation.

\begin{lstlisting}[
    style=quansical,
    language={},
    morekeywords={
        PLAN,QUANTUM,BACKEND,QULACS,
        SHOTS,RETURN,MODE,MEASUREMENT
    },
    keywordstyle=\bfseries,
    frame=single,
    breaklines=true,
    linewidth=\linewidth
]
qstore.snapshot_expectation(
    snapshot_id, state_id, bell_projector)
PLAN QUANTUM BACKEND QULACS SHOTS 4096
RETURN MODE MEASUREMENT;
\end{lstlisting}

In this QSQL, snapshot\_id selects the saved snapshot, and state\_id identifies the retained quantum state. snapshot\_expectation evaluates the Bell-state projector on this state to obtain its fidelity. PLAN QUANTUM requires a Quantum operator for this calculation, while BACKEND QULACS selects the simulator. SHOTS 4096 specifies the operation's sampling budget. RETURN MODE MEASUREMENT returns the fidelity estimate, sampling variance, and sample count.

\section{Hybrid Query Engine}
\label{sec:query}

\begin{figure}[!htbp]
  \centering
  \includegraphics[width=\columnwidth]{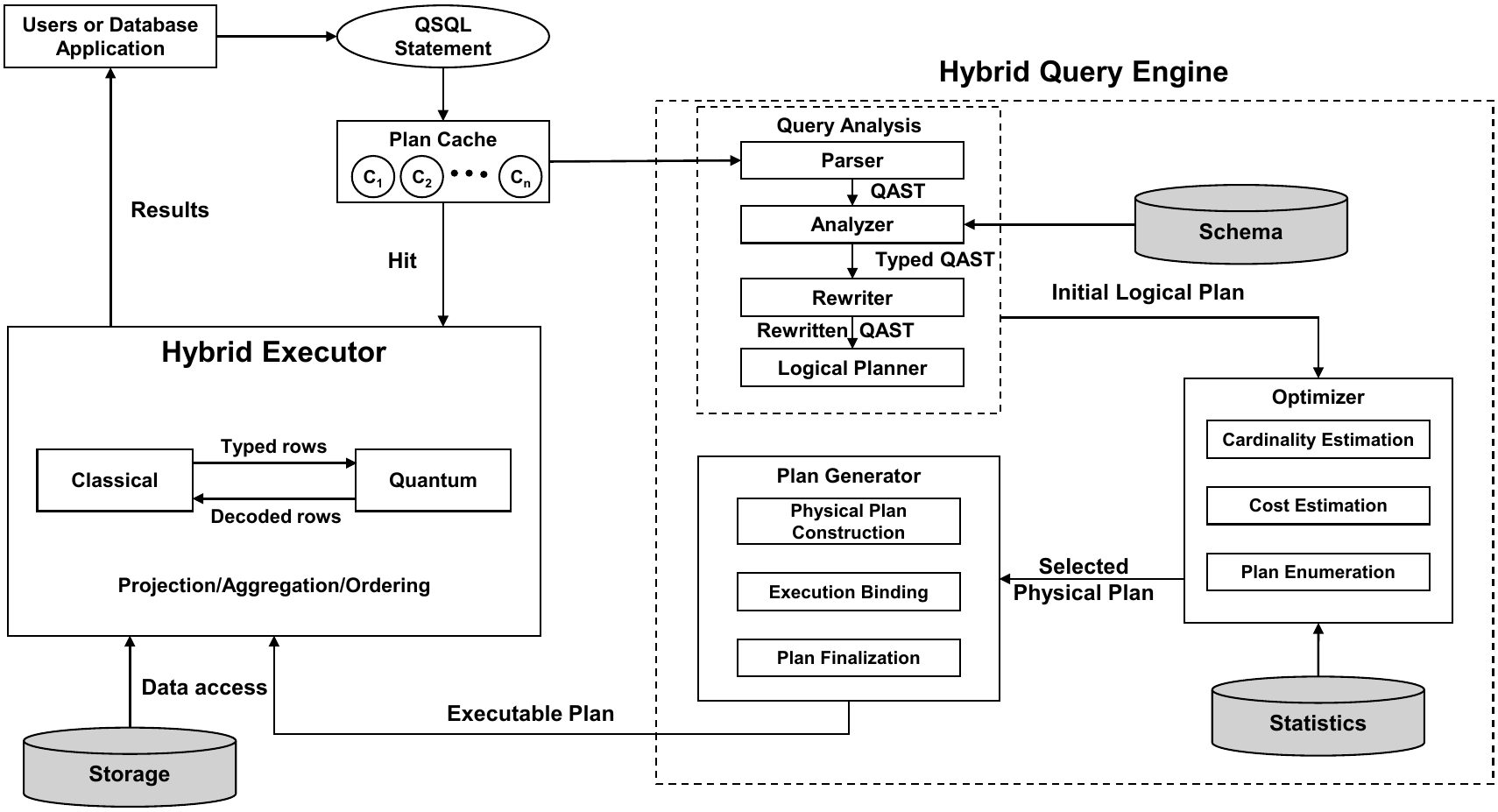}
  \caption{Hybrid Query Processing}
  \label{fig:hybrid-query}
\end{figure}

Considering the difficulties of quantum operator planning and cost estimation, \sys{} designs a Hybrid Query Engine to optimize classical and quantum operator combinations by total query cost, including quantum preparation and readout. Hybrid query processing comprises query analysis, hybrid optimization, and plan generation, as shown in Figure~\ref{fig:hybrid-query}.

\subsection{Query Analysis}

The parser parses QSQL statements into a Quantum Abstract Syntax Tree (QAST), which associates relational expressions with quantum execution requirements for hybrid planning. In the procedure of query analysis, semantic analysis resolves expression types and bindings, and  query rewriting normalizes the expressions and result requirements before the optimizer selects classical and quantum operators.

\textbf{Semantic Analysis.} The analyzer establishes expression types and bindings so that classical and quantum plans interpret QSQL consistently. The QAST extends a relational AST with quantum support annotations and execution requirements. Expression nodes record quantum support for functions and aggregates; query nodes separate execution policy from output form and result accuracy, so a query can require quantum execution with exact relational results. For approximate aggregation, the query node records the requested algorithm, measurement budget, and confidence level. For example, the following query requests amplitude estimation for the sum of \texttt{runs.energy}:
\begin{lstlisting}[basicstyle=\ttfamily\footnotesize,frame=none,numbers=none]
QSELECT SUM(energy) FROM runs
PLAN QUANTUM AGGREGATE MODE AMPLITUDE_ESTIMATION
SHOTS 1024 CONFIDENCE 0.95;
\end{lstlisting}
The analyzer binds \texttt{energy} to \texttt{runs.energy} and annotates \texttt{AMPLITUDE\_ESTIMATION} with its return type and quantum support. It combines these expression annotations with the execution policy and result requirements to produce the following QAST:
\begin{lstlisting}[basicstyle=\ttfamily\footnotesize,frame=none,numbers=none,keepspaces=true]
Query
  relation: runs
  target: AMPLITUDE_ESTIMATION
    argument: runs.energy : DOUBLE PRECISION
    result type: DOUBLE PRECISION
    quantum support: supported
  execution policy:
    intent: quantum required
    aggregate algorithm: amplitude estimation
    measurement budget: 1024
  result requirements:
    form: aggregate estimate
    accuracy: approximate
    confidence level: 0.95
\end{lstlisting}

\textbf{Query Rewriting.} The rewriter normalizes predicates in the QAST. The rewriter expands range predicates into lower and upper comparisons and simplifies Boolean expressions in the QAST, preserving SQL types and semantics. The rewritten QAST retains the execution policy, measurement budget, and result requirements and serves as input to the logical planner.

\textbf{Initial Logical Plan Construction.} The Logical Planner converts the rewritten QAST into a logical operator tree that connects relational processing with quantum state access and result readout. Each node records its inputs and references the QAST for expression types and bindings. Quantum state access specifies the inputs available to quantum operators. Measurement nodes describe observations of quantum states, and decoding nodes map measurement outcomes to relational tuples. For exact relational queries, the Logical Planner adds a completion node to require the recovery of all qualifying tuples. The logical planner associates the execution policy and result requirements with this operator tree to form the initial logical plan, then passes the plan and rewritten QAST to the optimizer.

\subsection{Hybrid Optimization}

\sys{} adopts hybrid query optimization to support QSQL's classical and quantum execution plan selection. It applies dynamic programming to enumerate and prune plans combining classical and quantum operators using cardinality estimates and a quantum cost model, then selects the lowest-cost candidate satisfying the execution requirements.

To estimate costs for plans combining classical and quantum operators, \sys{} introduces candidate cardinality. For an inner equijoin, the candidate cardinality $N_c$ counts tuple pairs satisfying the non-null join-key equalities, and the output cardinality $N_o$ counts pairs satisfying all join predicates. Together, they characterize the quantum search space and its qualifying fraction for quantum operator costing. The output cardinality also determines the input size of subsequent operators.

Our estimator derives the equality selectivity $s_{\mathrm{eq}}$ from join-key statistics, including distinct-value counts, NULL fractions, and most-common-value frequencies. The residual selectivity $s_{\mathrm{res}}$ estimates the fraction of candidate pairs satisfying the remaining join predicates. Given estimated input cardinalities $\widehat N_L$ and $\widehat N_R$, the estimator computes
\begin{equation}
\widehat N_c=\widehat N_L\widehat N_Rs_{\mathrm{eq}},
\qquad
\widehat N_o=\widehat N_cs_{\mathrm{res}}.
\end{equation}
The optimizer uses each subplan's estimated output cardinality as the input cardinality of its parent operator, allowing the cost model to account for intermediate result sizes when evaluating combinations of classical and quantum operators.

\textbf{Cost Estimation.}
To assess plans combining classical and quantum operators, \sys{} extends its cost model with quantum preparation, search, readout, and operator transition costs. The model considers the cost of intermediate-result materialization using estimated cardinalities and tuple widths. For input subplans $L$ and $R$, the total join cost is
\begin{equation}
\label{eq:join-cost}
C(L\Join_d R)=C(L)+C(R)+\Delta C_d(L,R),
\end{equation}
where $d\in\{\mathrm{c},\mathrm{q}\}$ denotes the classical or quantum join implementation. $C(L)$ and $C(R)$ include the costs of producing and materializing the inputs, while $\Delta C_d$ denotes the incremental join cost. For a quantum join,
\begin{equation}
\label{eq:quantum-join-cost}
\begin{aligned}
\Delta C_{\mathrm{q}}(L,R)
={}& C_{\mathrm{prepare}}+C_{\mathrm{search}}+C_{\mathrm{readout}}+C_{\mathrm{switch}}.
\end{aligned}
\end{equation}
$C_{\mathrm{prepare}}$ estimates join-key processing and quantum input preparation from the input cardinalities, join-key count, and candidate cardinality $\widehat N_c$. $C_{\mathrm{search}}$ estimates quantum search work using the candidate cardinality $\widehat N_c$, the qualifying output cardinality $\widehat N_o$, and the specified sampling and failure budgets. $C_{\mathrm{readout}}$ accounts for the required measurement, exact-result completion, and output construction using the measurement budget, $\widehat N_c$, output cardinality $\widehat N_o$, and output tuple width. $C_{\mathrm{switch}}$ refers to the cost of transitions between classical and quantum operators using a fixed transition cost. Each component converts its estimated work into cost using the corresponding unit cost.

\textbf{Plan Enumeration.}
\sys{} uses dynamic programming to enumerate join orders and classical/quantum operator combinations under the cost model. For each relation subset $S$ in the search space, it retains the cheapest classical subplan $P_{\mathrm{c}}(S)$ and the cheapest feasible subplan $P_{\mathrm{q}}(S)$ containing at least one quantum join.

The optimizer processes relation subsets in increasing order of size. For each eligible partition of $S$ into disjoint subsets $L$ and $R$, it combines retained child subplans with a classical or quantum join. It uses the cardinality estimates to evaluate the incremental join cost and adds the child costs according to Equation~\eqref{eq:join-cost}. A candidate containing a quantum join at the root or in either child competes for $P_{\mathrm{q}}(S)$, allowing classical and quantum joins at different positions in the same plan. The optimizer excludes candidates exceeding backend resource limits and updates $P_{\mathrm{c}}(S)$ or $P_{\mathrm{q}}(S)$ whenever it finds a cheaper feasible candidate in the corresponding category. For the complete query, automatic planning compares the classical and quantum-containing candidates and selects the feasible plan with the lowest estimated total cost.

\subsection{Plan Generation}

The plan generator resolves the state inputs and relational outputs of quantum computation, and converts the optimizer's selected classical and quantum operators into an executable plan with explicit state dependencies, quantum resource requirements, and readout operations.

\textbf{Physical Plan Construction.} The generator retains the selected join order when instantiating classical and quantum operators. A classical selection evaluates tuples directly, whereas a quantum selection searches candidate identifiers through an oracle derived from the query predicate. Value access supplies the oracle's inputs, and state preparation establishes the candidate state on which search operates. The generator connects these operations within the quantum subtree. When a classical parent consumes its output, decoding recovers the tuples identified by the selected candidates.

\textbf{Execution Binding.} The compiler associates the quantum subtree with the selected search procedure and binds its state and predicate inputs. Each candidate identifier denotes one tuple occurrence in the bound input, and the decoder uses this correspondence to retrieve SQL values while preserving duplicate rows. The procedure receives a register layout for the candidate address space and temporary predicate values, together with a sampling budget that bounds search trials. For approximate aggregation, the estimator also receives the confidence level used to compute the interval accompanying its result.

\textbf{Plan Finalization.} The compiler validates the generated program against the bound physical plan. Since quantum search may discover only part of the qualifying set, the verifier follows the search output to the enumeration or completion that produces an exact relational result. For approximate aggregation, the corresponding check follows the estimate through error computation. The bound register layout supplies the quantum workspace requirements checked against backend capacity before the program is passed to execution.

\section{Execution Engine}
\label{sec:execution}

\sys{}'s execution engine integrates database operations with quantum computation through a three-level instruction hierarchy. It compiles physical plans into executable instructions for the quantum backend. We also design two optimization strategies for the execution engine: Instruction Rewriting and Circuit Reduction, which reduce repeated computation and instruction dispatch and simplify quantum gate sequences, respectively.

\subsection{\sys{}-IR}
\label{sec:execution:hierarchy}

\begin{figure}[!htbp]
 \centering
 \includegraphics[width=\columnwidth]{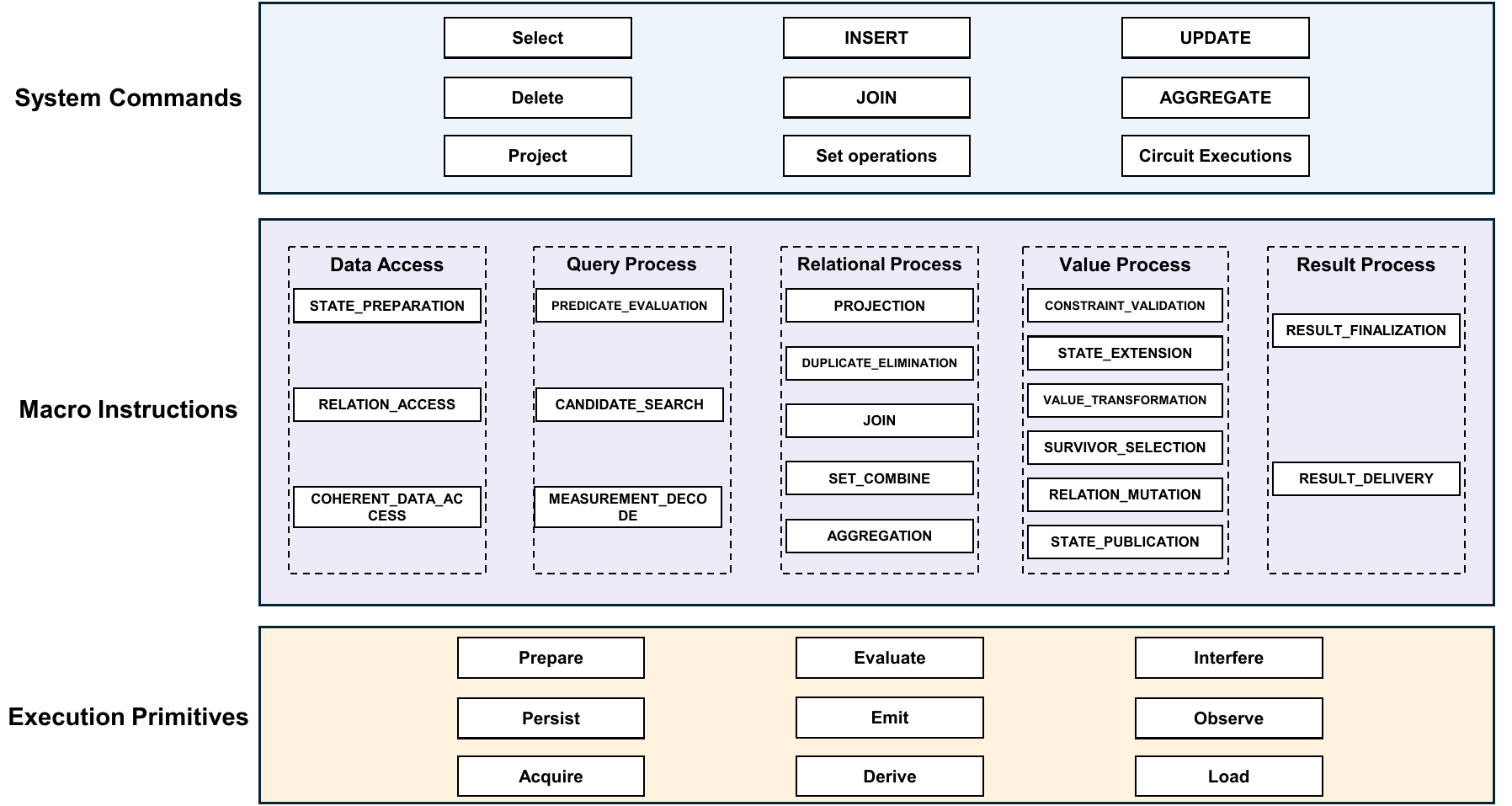}
 \caption{Three-Level Instruction Hierarchy of \sys{}-IR}
 \label{fig:instruction}
\end{figure}

Although existing studies propose the intermediate representation (IR) at different abstraction levels, including three-tier LLVM-style IR~\cite{qute} or QIR~\cite{stade2025qir}, to represent quantum operations through runtime interfaces, these IRs pay less attention to coordinating the quantum operations with database execution procedure. To compose database and quantum operations into complete execution programs and enable adaptation to various quantum backends through instruction interfaces, \sys{} introduces a novel three-level intermediate representation \sys{}-IR to its execution engine. Figure~\ref{fig:instruction} shows its structure.

Figure~\ref{fig:instruction} shows the three levels of \sys{}-IR: System Commands, Macro Instructions, and Execution Primitives. A system command specifies a database operation, such as aggregation or circuit execution. The compiler identifies the operation represented by the physical plan and constructs the system command from the plan's physical operators. For a Macro Instruction, it defines the execution procedure of database or quantum operations, and describes the computations required by physical or quantum operators. It implements the selected execution strategy by composing execution primitives, such as quantum search or aggregate estimation. Through Macro Instructions, System Commands can combine database processing with quantum computation and reuse these procedures across different database operations.

Execution Primitives define the basic operations of an execution procedure and specify the input data and results of each operation with the calling interface. The compiler translates these primitives into executable instructions for database and quantum processing backends, and each backend implements the corresponding operations through its own execution routines. The interpreter passes instruction inputs to these routines and supplies their results to subsequent instructions, combining database operations and quantum computation to complete QSQL processing.

The following QSQL statement illustrates our \sys{}-IR. It estimates the mean of the values in \texttt{oracle\_values} using maximum-likelihood amplitude estimation\cite{mlae}:
\begin{lstlisting}[style=quansical]
QSELECT AVG(value) AS estimate
FROM oracle_values
AGGREGATE MODE AMPLITUDE_ESTIMATION
PLAN QUANTUM BACKEND QULACS
SHOTS 4096 CONFIDENCE 0.95
RETURN MODE BOTH;
\end{lstlisting}

\sys{} compiles this statement into the System Command Aggregate and expands it according to the physical plan into the following Macro Instructions:
\begin{lstlisting}[style=quansical]
RelationAccess
CoherentDataAccess
PredicateEvaluation
Aggregation
ResultDelivery
\end{lstlisting}

The macro instruction RelationAccess uses the execution primitive Acquire to obtain the relation state under the statement snapshot, together with its row identifiers and encoding. The macro instructions CoherentDataAccess and PredicateEvaluation supply the encoded values and predicate through the Execution Primitives Load and Evaluate. These outputs become operands of the macro instruction Aggregation. The corresponding expansion is:
\begin{lstlisting}[style=quansical]
RelationAccess      : Acquire
CoherentDataAccess  : Load, Load (adjoint)
PredicateEvaluation : Evaluate, Evaluate (adjoint)
Aggregation         : Interfere, Observe, Derive
ResultDelivery      : Emit
\end{lstlisting}

Within the Aggregation macro instruction, the execution primitive Interfere represents controlled rotations on normalized values and amplitude amplification. The execution primitive Observe supplies measurement counts to the execution primitive Derive for maximum-likelihood estimation and conversion back to the original value scale. The adjoint forms undo temporary predicate and value preparation. The interpreter executes amplification, observation, and estimation in one backend call using the requested shot budget and confidence level. The execution primitive Emit returns the estimate, confidence interval, and sampling metadata.

\subsection{Hybrid Execution}
\label{sec:execution:runtime}

The execution engine dispatches Execution Primitives to database and quantum backends according to their execution domains. The database backend evaluates classical relational operators over tuples. The quantum backend implements the quantum operators selected by the physical plan. 

The execution engine accesses relation inputs and stored quantum states through the Storage Engine under the statement snapshot. For a stored quantum input, the Storage Engine resolves the visible relation version and retrieves its amplitudes and encoding. The associated row mapping connects quantum basis positions to logical row identifiers and visible tuples. When an operator accesses a subset of rows, the Storage Engine extracts and normalizes the corresponding amplitudes. The execution engine supplies the resulting state and required column values to the quantum procedure. For newly prepared inputs, the quantum backend constructs the state from the accessed tuples and their encoding.

Backend routines return their results to the interpreter through the execution interface. The execution engine decodes quantum candidate identifiers into visible tuples and invokes classical completion when specified by the plan. Classical estimation converts quantum measurement counts into aggregate values and confidence intervals. These tuples and aggregate values enter subsequent relational operators according to the physical plan. The execution engine coordinates classical and quantum processing through instruction dependencies, and Emit returns the final QSQL result.

\subsection{Execution Optimization}
\label{sec:execution:optimization}

Circuit depth and gate duration contribute directly to quantum query latency~\cite{qute}. In quantum circuit processing, repeated gate application can make memory access the execution bottleneck~\cite{qulacs}. \sys{} introduces Instruction Rewriting and Circuit Reduction to reduce instruction processing and quantum gate operations.

Instruction Rewriting applies common subexpression elimination to repeated data preparation and predicate evaluation. The rewriter identifies equivalent computations from their operations, operands, and parameters while tracking updates to the accessed state. For shared reversible preparation, it retains one adjoint operation after the final use. During circuit compilation, the compiler represents the gate sequence as one circuit block executed by the Interfere primitive. It constructs this block directly during instruction generation, preserving gate order while reducing instruction construction and interpreter dispatch.

Circuit Reduction simplifies the gate sequence in circuit execution. The optimizer uses dependencies between gate targets and controls to cancel inverse pairs and merge rotations about the same axis. It further combines successive single-qubit operations by multiplying their matrices in execution order. Measurements and other nonunitary operations delimit the sequences eligible for these transformations. Circuit Reduction precedes instruction generation, so Instruction Rewriting constructs a circuit block from the simplified sequence. The backend adapter compiles the same sequence for execution.

We use the following circuit to illustrate our execution optimization strategies. It applies eight gates to two qubits before result readout:
\begin{lstlisting}[style=quansical]
h q[0];  x q[1];  h q[0];
ry(pi/8) q[0];  ry(pi/8) q[0];
cx q[1], q[0];  cx q[1], q[0];
cx q[0], q[1];
\end{lstlisting}

Circuit Reduction cancels the two Hadamard gates across the independent X gate, merges the rotations into one rotation by $\pi/4$, and cancels the identical controlled-X pair. The resulting sequence contains three gates:
\begin{lstlisting}[style=quansical]
x q[1];
ry(pi/4) q[0];
cx q[0], q[1];
\end{lstlisting}

Without either optimization, the complete operation requires eleven execution instructions: eight for the gates and three for state preparation, result readout, and result delivery. Instruction Rewriting represents the eight gates with one instruction for the Execution Primitive Interfere, reducing the total to four. Circuit Reduction alone reduces the gate instructions from eight to three, leaving six instructions. Together, the strategies produce four instructions containing three gates. The database execution engine thus constructs and processes fewer instructions, while the quantum backend applies fewer gates.

\section{Storage Engine}
\label{sec:state}
\sys{} designs a storage engine, with version management, state reconstruction, and cost-based state reuse, to reduce repeated amplitude storage. By setting shared checkpoints and using operation histories, our storage engine can reduce the cost of state persistent storage and circuit execution while preserving the quantum data version associations. 

\subsection{Version Management}
\label{sec:state-versioning}

Tuple versions record row changes but do not capture how quantum states are derived, making the reconstruction of quantum state difficult. \sys{} therefore defines a data version, which associates each quantum data version with its input states, operations, and source tuple versions. It also introduces the database version that retains references to selected quantum data versions and tuple versions, together with their derivation histories, for historical queries and reuse of intermediate quantum results across transactions.

A quantum data version comprises a quantum state identified by a UUID and records describing its derivation. An operation record identifies the input quantum data versions and the applied circuit, especially for a tensor operation, it identifies both inputs and their qubit positions in the output.  For states prepared from relational data, a qindex maps basis positions to logical row identifiers and the physical tuple identifiers (TIDs) of their source versions. Each entry contains a logical row identifier, which identifies the row across updates, and a tuple identifier (TID), which locates its physical row version. \sys{} represents a quantum database version as a persistent snapshot that retains selected quantum states and relational tuples together with the states' computational histories. 

\sys{} organizes quantum data version storage around checkpoints. A checkpoint stores the amplitudes of a quantum data version under its UUID. \sys{} creates checkpoints when operation depth reaches a configurable threshold or upon an explicit request. Between checkpoints, versions are represented by operation records referencing earlier inputs. Branches and database snapshots share these checkpoints and their histories. Snapshot retention follows these records to the input states and uses their qindexes to identify the source tuples to retain. When a query accesses a version without stored amplitudes, \sys{} traces its operation history to an available checkpoint and replays the intervening operations to reconstruct the requested state. This sharing reduces repeated storage of common inputs and repeated execution of common circuit prefixes.

\begin{figure}[!htbp]
  \centering
  \includegraphics[width=\columnwidth]{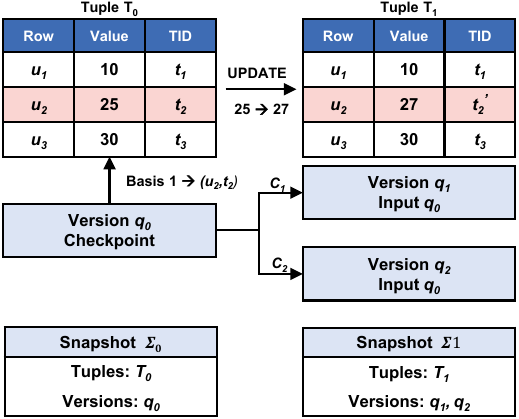}
  \caption{Quantum data versions and persistent database snapshots.}
  \label{fig:quantum-version-management}
\end{figure}

Figure~\ref{fig:quantum-version-management} illustrates quantum data version management with a tuple update and subsequent quantum computations. Initially, quantum data version $q_0$ is prepared from tuple set $T_0$, and database snapshot $\Sigma_0$ retains both. Updating row $u_2$ from 25 to 27 creates a new tuple version $t'_2$ in $T_1$. Meanwhile, applying circuits $C_1$ and $C_2$ to $q_0$ produces quantum data versions $q_1$ and $q_2$, both referencing the shared input $q_0$. Database snapshot $\Sigma_1$ retains $T_1$, $q_1$, and $q_2$. A relational query on $\Sigma_1$ reads the updated value 27, whereas tracing the input references of $q_1$ and $q_2$ to $q_0$'s qindex identifies the original tuple $t_2$ containing 25 at basis position 1.

\subsection{Historical State Reconstruction}
\label{sec:state-reconstruction}

Quantum data reconstruction recovers the complex amplitudes of historical states so that database queries can reuse intermediate results from earlier computations. \sys{} uses the version history to support forward replay from shared checkpoints and inverse execution from retained successors. Both paths reconstruct the quantum data at the target version.

\begin{figure}[!htbp]
  \centering
  \includegraphics[width=\columnwidth]{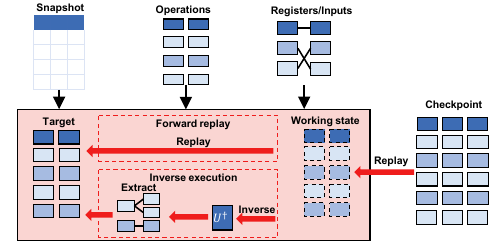}
  \caption{Historical State Reconstruction through Forward Replay and Inverse Execution}
  \label{fig:reconstruction-strategy}
\end{figure}

Figure~\ref{fig:reconstruction-strategy} shows the two reconstruction paths. Forward replay follows the target's input references to a visible checkpoint or decoded cached state, then executes the intervening operations in computation order. Inverse execution starts from a retained successor connected to the target by reversible operations. It applies their inverses in reverse order to a working copy of the successor's state vector, using the recorded register mapping. If the successor is not materialized, forward replay reconstructs it before inverse execution. The reconstructed quantum state supports historical query evaluation and serves as input for deriving new state versions.

\begin{figure}[!htbp]
  \centering
  \includegraphics[width=\columnwidth]{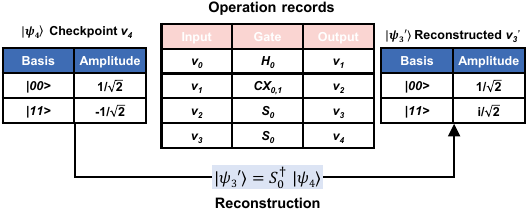}
  \caption{Example of Inverse State Reconstruction from a Successor Checkpoint}
  \label{fig:state-reconstruction}
\end{figure}

Figure~\ref{fig:state-reconstruction} illustrates inverse execution for recovering the target state $v_3=(|00\rangle+i|11\rangle)/\sqrt{2}$ from a successor checkpoint $v_4=(|00\rangle-|11\rangle)/\sqrt{2}$. Applying the recorded inverse gate $S_0^\dagger$ to $v_4$ restores the relative phase of the $|11\rangle$ amplitude, recovering $v_3$. The two versions have identical computational-basis probabilities. Reconstruction thus restores the phase of $v_3$ for historical query processing.

The two paths also cooperate when the supplied successor requires reconstruction. Forward replay obtains that successor from an available checkpoint or cached state, and inverse execution follows its recorded history back to the target. A checkpoint can therefore serve later versions through replay and earlier versions through inverse execution along a reversible history. \sys{} stores the intervening derivations as operation records, allowing queries on different versions to share the checkpoint's amplitude storage and reconstruct their requested states on demand.

\subsection{Cost-Based Quantum Data Reuse}
\label{sec:state-reuse}

Quantum state reuse accelerates query processing using states produced by earlier computations. Building on version management and historical reconstruction, \sys{} chooses the better strategy between inverse reconstruction and observable propagation estimated costs for historical queries.

For a target state $|\psi\rangle$ and retained successor $|\phi\rangle=U|\psi\rangle$, where $U$ is the recorded unitary evolution, inverse reconstruction recovers the target as $U^\dagger|\phi\rangle$ and evaluates the requested observable $O$.  Observable propagation instead evaluates the transformed observable $O'=UOU^\dagger$ on the retained successor $|\phi\rangle$:
\begin{equation}
\langle\psi|O|\psi\rangle
=\langle\phi|UOU^\dagger|\phi\rangle.
\label{eq:historical-observable}
\end{equation}
\sys{} constructs $O'$ by propagating the Pauli terms of $O$ through the recorded gates and combining identical terms.

\sys{} estimates the processing costs of both strategies using state-vector size, recorded gate operations, and Pauli-term counts. For inverse reconstruction, the cost includes copying the successor state, executing inverse gates, and evaluating $O$ on the recovered state. For observable propagation, the cost includes transforming and combining Pauli terms through the recorded operations and evaluating $O'$ on the retained successor. The estimator accounts for amplitude accesses during state operations and observable evaluation, as well as the Pauli terms processed during propagation. Since propagation may increase the number of Pauli terms, its transformation and evaluation costs can outweigh the cost of inverse reconstruction. Thus, \sys{} selects observable propagation when avoiding inverse execution provides greater savings, and inverse reconstruction when observable transformation becomes more expensive.

\section{Evaluation}
\label{sec:evaluation}
\label{sec:experiments}
We conduct extensive experiments to evaluate \sys{} under different workloads and compare it with various classical relational databases and quantum databases.

\subsection{Experimental Setup}
\noindent\textbf{Hardware settings.} We deploy \sys{} on a physical server running Ubuntu, equipped with an AMD Ryzen 9 9950X3D processor @ 4.3 GHz, 256 GB of RAM, and 4 TB of disk storage. The Linux kernel version is 6.8.0-136-generic().

\noindent\textbf{Implementation.} We implement \sys{} on the open-source system. We implement the hybrid query engine, execution engine, and storage engine based on PostgreSQL 18.4 using default settings. We use Qulacs~\cite{qulacs} without any tuning as the quantum processing backend. We set the PLAN AUTO as the default operator selection mode of \sys{}.

\noindent\textbf{Baseline systems \& methodologies.}
For relational query processing, we use PostgreSQL 18.4, SQLite 3.46.1, Qute~\cite{qute}, and DEDALUS~\cite{dedalus}. We adapt Qute to PostgreSQL 18.4 and use Qulacs~\cite{qulacs} for quantum sampling in Qute and DEDALUS.  Moreover, for quantum data processing, these baselines use the same backend Qulacs and compiled user-defined functions (UDFs) for quantum operations and dense state vectors for storage.  For circuit execution, we use Qiskit Aer 0.17.0 and Qulacs, and extend PostgreSQL and SQLite with InfiniQuantumSim's SQL translator to execute circuits~\cite{inferq}. For version management, we evaluate three other methods in \sys{}: Full Snapshot~\cite{elnozahy2002rollback} stores a complete state per version; Event Sourcing~\cite{overeem2021eventsourcing} stores initial checkpoints and operation histories; Incremental Checkpointing~\cite{plank1995libckpt} stores changed amplitude blocks. For state reconstruction, we also assess Full Recomputation~\cite{qtask}, which executes each circuit from scratch, and Event Sourcing which reconstructs and caches the shared-prefix state after the first task. For historical queries, we compare three other strategies: Fixed Inverse~\cite{luo2020yao} reconstructs the target state from a successor; Fixed Propagation~\cite{rall} evaluates propagated observables on that successor; Forward Replay~\cite{elnozahy2002rollback} reconstructs the target state from an ancestor checkpoint.


\subsection{Performance of Query Processing}
\label{PQP}
\textbf{Performance of relational query processing.} We evaluate the relational query processing performance of \sys{} using queries in STATS-CEB on the STATS dataset, which contains eight tables and 1,029,842 rows~\cite{statsceb}. The query set consists of 33 queries: eight low-complexity queries join 2--3 relations, and 25 high-complexity queries join 3--6. All systems run in the same hardware environment and execute identical queries. We measure query latency, execution time, and query compilation time.

\begin{table*}[!htbp]
\centering
\caption{Query Latency, Execution Time, and Compilation Time Comparison on STATS}
\label{tab:query}
\fontsize{8}{9.6}\selectfont
\setlength{\tabcolsep}{3pt}
\renewcommand{\arraystretch}{1.12}
\begin{tabular*}{\textwidth}{@{\extracolsep{\fill}}c|ccc|ccc|ccc}
\hline
\multirow{2}{*}{System}
& \multicolumn{3}{c|}{Query Latency (ms)}
& \multicolumn{3}{c|}{Execution Time (ms)}
& \multicolumn{3}{c}{Query Compilation Time (ms)} \\
\cline{2-10}
& \shortstack{Low\\Complexity} & \shortstack{High\\Complexity} & Total
& \shortstack{Low\\Complexity} & \shortstack{High\\Complexity} & Total
& \shortstack{Low\\Complexity} & \shortstack{High\\Complexity} & Total \\
\hline
Ours & \textbf{957.074} & \textbf{1820.779} & \textbf{2777.853}
& \textbf{931.785} & \textbf{1730.631} & \textbf{2662.416}
& 25.289 & 90.148 & 115.437 \\
\hline
SQLite & 33675.980 & 3915.375 & 37591.354
& 33674.902 & 3911.932 & 37586.833
& \textbf{1.078} & \textbf{3.443} & \textbf{4.521} \\
\hline
PostgreSQL & 2548.472 & 3389.296 & 5937.768
& 2538.789 & 3359.329 & 5898.117
& 9.683 & 29.967 & 39.650 \\
\hline
Qute & 2536.572 & 4305.121 & 6841.692
& 2526.560 & 4273.533 & 6800.093
& 10.011 & 31.588 & 41.599 \\
\hline
DEDALUS & 2696.493 & 6221.215 & 8917.708
& 2541.683 & 5467.359 & 8009.042
& 154.810 & 753.856 & 908.666 \\
\hline
\end{tabular*}
\par\smallskip
\end{table*}

\begin{figure*}[!htbp]
\centering
\includegraphics[width=\textwidth]{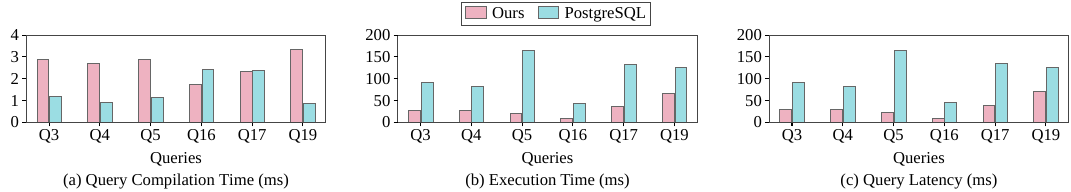}
\setlength{\textfloatsep}{0pt}
\caption{Query Compilation Time, Execution Time, and Latency Comparison on STATS}
\label{fig:evaluation-queries}
\end{figure*}

Table~\ref{tab:query} shows query latency, execution time, and compilation time across the five systems on the query sets. \sys{} achieves the lowest latency in both complexity groups, with an overall $2.14\times$ speedup over the fastest baseline, PostgreSQL, and $2.46$--$13.53\times$ over the others. The speedups over PostgreSQL are $2.66\times$ for low-complexity queries and $1.86\times$ for high-complexity queries.

The benefits are derived from the selection of better executable plans. Although analyzing quantum requirements and selecting operator combinations increase query compilation time, the selected plans reduce execution time in the Execution Engine. The reduction in query latency shows that these execution savings compensate for the additional compilation cost. Figure~\ref{fig:evaluation-queries} further shows the trade-off. Although \sys{} spends more time on query compilation than PostgreSQL across 4 queries (Figure~\ref{fig:evaluation-queries}(a)), the selected plans reduce execution time (Figure~\ref{fig:evaluation-queries}(b)), and these savings lower overall query latency (Figure~\ref{fig:evaluation-queries}(c)).

\textbf{Benefits of Hybrid Operator Selection.} To evaluate the effectiveness of hybrid operator selection, we compare QuarkDB in AUTO and CLASSICAL modes on Q16 and Q17. The CLASSICAL mode restricts plan selection to classical operators, whereas AUTO jointly considers classical and quantum operators. Table\ref{tab:query-phases} reports the functional-stage execution times under both configurations, together with PostgreSQL using the default settings.

\begin{table}[!htbp]
\centering
\caption{Query Processing Time Breakdown for Q16 and Q17 (ms)}
\label{tab:query-phases}
\fontsize{6.5pt}{7.8pt}\selectfont
\setlength{\tabcolsep}{1.5pt}
\renewcommand{\arraystretch}{1.12}
\begin{tabular*}{\columnwidth}{
    @{\extracolsep{\fill}}ccccccc@{}}
\toprule
& \multicolumn{3}{c}{\textbf{Q16}}
& \multicolumn{3}{c}{\textbf{Q17}} \\
\cmidrule(lr){2-4}
\cmidrule(l){5-7}
\textbf{Stage}
& \textbf{PG}
& \textbf{Auto}
& \textbf{Classical}
& \textbf{PG}
& \textbf{AUTO}
& \textbf{CLASSICAL} \\
\midrule
Preparation
& 0.088 & 0.788 & 0.625
& 0.115 & 0.649 & 0.531 \\
Optimization
& 0.500 & 2.104 & 0.750
& 0.580 & 7.751 & 0.959 \\
Plan generation
& 0.007 & 0.230 & 0.016
& 0.012 & 0.157 & 0.022 \\
\addlinespace[0.2em]
Initialization
& 0.012 & 0.067 & 0.011
& 0.019 & 0.091 & 0.343 \\
Scan/filter
& 8.092 & 4.503 & 11.813
& 27.519 & 26.873 & 45.920 \\
Join
& 11.002 & \textbf{2.114} & 10.937
& 130.761 & \textbf{3.588} & 9.954 \\
Aggregation
& 0.009 & 0.090 & 0.010
& 5.265 & 0.308 & 55.243 \\
Output
& 0.004 & 0.002 & 0.002
& 0.005 & 0.002 & 0.006 \\
Control
& 0.152 & 0.188 & 0.198
& 0.111 & 0.610 & 0.280 \\
Probe bookkeeping
& 0.015 & 0.008 & 0.013
& 1.002 & 0.077 & 3.264 \\
Client residual
& 0.028 & 0.026 & 0.029
& 0.204 & 0.035 & 0.058 \\
\midrule
\textbf{Total latency}
& 19.910 & \textbf{10.121} & 24.404
& 165.592 & \textbf{40.141} & 116.578 \\
\bottomrule
\end{tabular*}
\par\smallskip
    \parbox{\columnwidth}{PG: PostgreSQL; AUTO and CLASSICAL: \sys{} Execution PLAN.}
\end{table}

Compared with the CLASSICAL mode, AUTO achieves $2.41\times$ and $2.90\times$ speedups in end-to-end query latency on Q16 and Q17, respectively. In particular, during execution the selected hybrid plans achieve 5.17$\times$ and 2.77$\times$ speedups in join processing on Q16 and Q17, respectively. The hybrid plans also accelerate scan/filter processing on both queries and aggregation on Q17. Compared with PostgreSQL, AUTO achieves 1.97$\times$ and 4.13$\times$ end-to-end speedups on Q16 and Q17, respectively. The results illustrate that cost-based hybrid plan selection introduces additional optimization overhead, but the execution time savings outweigh this overhead, resulting in lower overall query latency.

\textbf{Quantum Data Processing.} Moreover, we use NetSquid repeater-chain data~\cite{labaymora2024,entanglementdata2023} to evaluate the performance of quantum data processing. The dataset contains two-qubit density matrices with protocol and delivery-time metadata. The experiment selects 60 records and combines every five independent entangled pairs into one joint state, producing 12 states. Each state uses ten physical qubits and ten ancillary qubits to represent the mixed states. The workload covers state construction through query-result retrieval, which constructs and stores these joint states and computes mean Bell-pair fidelity by protocol. We measure the complete latency to assess the query processing performance, preparation time to analyze the circuit preparation cost, commit time to evaluate state storage cost and fidelity latency to calculate fidelity-analysis cost. 

\begin{figure*}[!htbp]
\centering
\includegraphics[width=\textwidth]{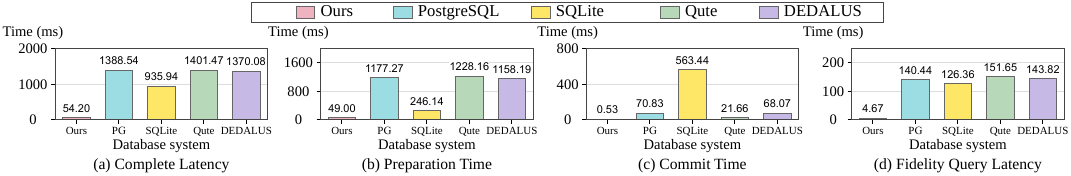}
\caption{Latency Comparison for Quantum Data Processing}
\label{fig:evaluation-quantum-data}
\end{figure*}

Figure~\ref{fig:evaluation-quantum-data} shows that \sys{} outperforms all baselines, achieving speedups of $17.27$--$25.86\times$ in complete latency and even $27.06$--$32.47\times$ in fidelity-query latency over the four baselines. These improvements arise because our system integrates quantum operations into the entire database execution process, thereby reducing the significant data encoding and access cost of processing quantum data using traditional relational databases. While some quantum databases like Qute or DEDALUS introduce quantum functionality into query processing, they still encode quantum data in a traditional manner, resulting in poorer performance.

\subsection{Performance of \sys{} Execution}
\label{PCE}

\textbf{Quantum Circuit Execution Performance.} We evaluate the circuit compilation and execution performance of \sys{} using the InferQ dataset, which contains 202,975 quantum circuits assembled from circuit templates~\cite{inferq}. We select circuits at different qubit widths, and each $n$-qubit circuit contains $2n$ RY gates, $2n$ RZ gates, and $n-1$ CNOT gates. All systems run in the same hardware environment and execute identical circuits. We measure circuit execution time and end-to-end latency to assess quantum execution performance and overall response time, respectively. 

\begin{figure}[!htbp]
\centering
\includegraphics[width=\columnwidth]{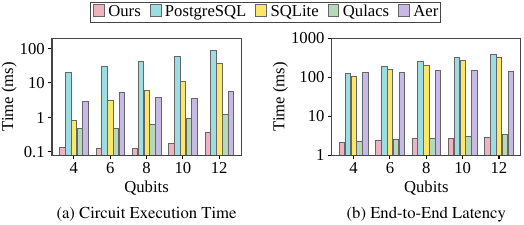}
\setlength{\abovecaptionskip}{2pt}
\caption{Circuit Execution Time and End-to-End Latency Comparison on InferQ (Log Scale)}
\label{fig:evaluation-circuits}
\end{figure}

Figure~\ref{fig:evaluation-circuits} illustrates that \sys{} achieves $3.25$--$5.30\times$ execution speedups and $1.04$--$1.16\times$ end-to-end latency speedups over Qulacs, the fastest baseline. Our system outperforms other database because \sys{}-IR reduces amplitude updates and instruction dispatch by gate simplification and fusion. For example, at 12 qubits circuits simplification reduces 59 input gates to 35 logical gates, and fusion produces five backend operations dispatched through four instructions, thereby reducing latency. 

\textbf{Ablation Study.} To understand the benefits of Instruction Rewriting and Circuit Reduction described in Section~\ref{sec:execution}, we compare the system performance under ablation configurations: Neither, Instruction Rewriting, Circuit Reduction, and Both. 
Neither and Both disable and enable both optimizations, respectively, while Instruction Rewriting and Circuit Reduction enable only their corresponding optimizations.

\begin{table}[!htbp]
\centering
\caption{Ablation Study of Instruction Rewriting and Circuit Reduction}
\label{tab:ablation}
\fontsize{8}{9.6}\selectfont
\setlength{\tabcolsep}{2pt}
\renewcommand{\arraystretch}{1.16}
\begin{tabular*}{\columnwidth}{@{\extracolsep{\fill}}cccccc@{}}
\toprule
\multirow{2}{*}{Configuration} & \multicolumn{5}{c}{Quantum Backend Processing Time (ms)} \\
\cmidrule(lr){2-6}
 & 4 qubits & 6 qubits & 8 qubits & 10 qubits & 12 qubits \\
\midrule
Neither & 2.892 & 0.729 & 2.113 & 1.047 & 2.344 \\
Instruction Rewriting & 0.557 & 0.691 & 0.713 & 0.849 & 1.307 \\
Circuit Reduction & 1.478 & 0.597 & 1.194 & 0.998 & 1.973 \\
Both & \textbf{0.521} & \textbf{0.472} & \textbf{0.508} & \textbf{0.769} & \textbf{1.040} \\
\bottomrule
\end{tabular*}
\end{table}

Table~\ref{tab:ablation} shows \sys{} achieves the largest speedup with Both at every qubit width, ranging from $1.36\times$ to $5.55\times$. At 8 qubits, Instruction Rewriting reduces 147 instructions to four while retaining 144 gates; Circuit Reduction reduces the gate count to 62. Combining them produces 62 gates in four instructions and a $4.16\times$ speedup. The two transformations remove different work: Circuit Reduction simplifies quantum operations, while Instruction Rewriting avoids repeated instruction construction, validation, and binding. The varying speedups across widths reflect these differences in instruction repetition and circuit structure. 

\textbf{Effectiveness of historical data reuse.} We use the 20-qubit Ising evolution~\cite{lamb2024ising} to evaluate the benefits of reusing historical data in queries. The workload retains 38 state versions, and five queries compute local and mean magnetization at three recorded times. We compare the performance of \sys{} under four methods for processing historical queries: cost-based reuse, Fixed Inverse, Fixed Propagation, and Forward Replay. The evaluation measures complete workload latency and per-query latency to assess historical query performance.

\begin{table}[!htbp]
\centering
\caption{Historical Query Workload Latency Comparison}
\label{tab:historical-queries}
\tablesize
\setlength{\tabcolsep}{4pt}
\renewcommand{\arraystretch}{1.10}
\begin{tabular*}{\columnwidth}{@{\extracolsep{\fill}}ccc@{}}
\toprule
Method & Latency (ms) & Speedup \\
\midrule
Ours & \textbf{213.27} & $\mathbf{9.63\times}$ \\
Fixed Inverse & 260.17 & $7.89\times$ \\
Fixed Propagation & 281.58 & $7.29\times$ \\
Forward Replay & 2053.83 & $1.00\times$ \\
\bottomrule
\end{tabular*}
\par\smallskip
    \parbox{\columnwidth}{Speedup is relative to Forward Replay.}
\end{table}

Table~\ref{tab:historical-queries} shows that \sys{} achieves the lowest latency. The cost-based reuse strategy selects inverse reconstruction or observable propagation according to each query's estimated execution cost. It evaluates propagated local observables on retained successor states to save reconstruction work. For mean magnetization, it selects inverse reconstruction to reduce the cost of evaluating an expanded observable. Reuse of successor states also reduces replay from ancestor checkpoints. These reductions yield complete-workload speedups of $1.22\times$, $1.32\times$, and $9.63\times$ over Fixed Inverse, Fixed Propagation, and Forward Replay, respectively.

\subsection{Performance of Version Management and State Reconstruction}
\label{PSR}
We evaluate the benefits of version management and state reconstruction using a 16-qubit QAOA maximum-cut circuit workload~\cite{qaoa} on a three-regular graph with 16 vertices and 24 edges. The workload performs 256 reads and 256 one-layer updates on four quantum states and retains all 260 versions to evaluate version management overhead.  

We measure reconstruction latency, including database restart and retrieval of the four committed states, and WAL volume to analyze the write cost of retaining these versions. To evaluate the reconstruction correctness, we assess reconstruction fidelity against independently computed reference states using $F_v=|\langle\psi_v\mid\hat{\psi}_v\rangle|^2$ for normalized states, with an error threshold of $1-F_v\le10^{-12}$. Moreover, we evaluate the benefits of reconstructed states for task processing. The workload further executes batches of 8, 16, or 32 tasks with an eight-layer shared prefix and a two-layer suffix specific to each task. We use batch latency and circuit time to measure the savings from supplying subsequent tasks with the reconstructed prefix state. 

\begin{figure}[!htbp]
\centering
\includegraphics[width=\columnwidth]{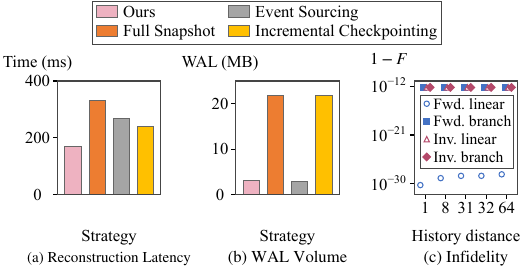}
\caption{State Reconstruction Latency, WAL Volume, and Infidelity Comparison}
\label{fig:evaluation-reconstruction}
\end{figure}

Figure~\ref{fig:evaluation-reconstruction} shows that \sys{} reduces reconstruction latency by 30.0\% compared with the best baseline Event Sourcing. \sys{} retains all 260 versions using operation histories and eight checkpoint versions sharing two distinct state payloads. While additional writes generate slightly more WAL than Event Sourcing, retained checkpoints of \sys{} provide starting states closer to the committed version, resulting in the lower latency.  Full Snapshot stores every checkpoint, and Incremental Checkpointing requires the same number checkpoint because QAOA updates modify every amplitude block. Consequently, \sys{} generates 85.3\% less WAL than these two methods (Figure~\ref{fig:evaluation-reconstruction}(b)). Figure~\ref{fig:evaluation-reconstruction}(c) shows the infidelity of \sys{}, covering 40 forward and inverse reconstructions at distances of 1, 8, 31, 32, and 64 circuit operations under linear and branched histories of, before and after restart. The maximum infidelity is $6.60\times10^{-13}$, which is below the corresponding threshold. And corresponding results before and after restart coincide, confirming that our state reconstruction can accurately generate the intended quantum state.

\begin{figure}[!htbp]
\centering
\includegraphics[width=\columnwidth]{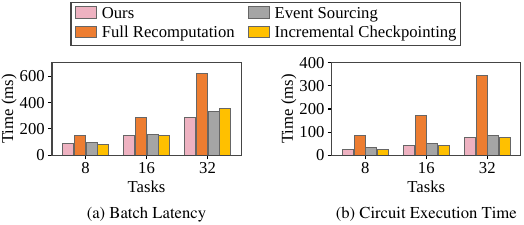}
\caption{Batch Latency and Circuit Execution Time Comparison for Shared-Prefix Tasks}
\label{fig:evaluation-reuse}
\end{figure}

Moreover, Figure~\ref{fig:evaluation-reuse}(a) shows that \sys{} achieves the lowest batch latency at 32 tasks, reducing it by 13.7\% compared with Event Sourcing. \sys{} loads the retained checkpoint of the shared prefix to supply inputs to subsequent tasks, while Event Sourcing replays this prefix once after the first task commits and then caches it. Compared with Full Recomputation, the batch-latency speedup increases from $1.66\times$ at eight tasks to $2.17\times$ at 32 tasks. The shared eight-layer prefix executes once, and each task executes only its two-layer suffix. At 32 tasks, this reduces executed layers from 320 to 72 and yields a $4.56\times$ circuit speedup. 
\section{Related Work}
\label{sec:related}

\subsection{Quantum Data Management}
Quantum data management addresses the representation and processing of quantum states. Hai et al.~\cite{hai2025} discuss data models, transactions, and recovery for quantum applications. Rieger et al.~\cite{rieger} define preparation, writing, readout, and deletion operations over index and data registers. Li et al.~\cite{li2024} propose quantum row-oriented and column-oriented table representations and analyze their qubit and gate requirements. Qymera~\cite{qymera} translates quantum circuits into SQL queries, representing amplitudes and gate matrices as tables. \sys{} represents relation versions, quantum states, circuits,
and classical results through a unified object model, and provides QSQL supporting quantum query processing.

\subsection{Hybrid Query Optimization}
Quantum approaches encode multiple query optimization~\cite{trummer2016} and join ordering~\cite{schoenberger2023join} as quadratic unconstrained binary optimization (QUBO) problems. QDBO~\cite{qdbo} reduces QUBO instances for quantum annealing and corrects them using sampling feedback. DEDALUS~\cite{dedalus} weights join subsets using database statistics and prunes disconnected subsets before QUBO construction. Qute~\cite{qute} compares classical and quantum operators by latency, success probability, and approximation error, using selective indexing to reduce encoded data. \sys{} uses dynamic programming to select execution plans with classical and quantum operators based on estimated total cost, retaining the lowest-cost classical and quantum-containing subplans for each relation subset. Its cost model distinguishes candidate and output cardinalities and accounts for input preparation, quantum search, result readout, and transitions between classical and quantum operators.

\subsection{Quantum Compilation and Execution}
Quantum compilers use intermediate representations for optimization before hardware-specific translation~\cite{quantumcompilation}. ProjectQ~\cite{projectq} composes decomposition, local optimization, and execution engines. QIRO~\cite{qiro} exposes quantum and classical data dependencies for optimization; Quartz~\cite{quartz} optimizes circuits through generated and verified transformations. \sys{} integrates database operations with quantum execution through three-level instructions intermediate representation \sys{}-IR, System Commands, Macro Instructions, and Execution Primitives. It introduces Instruction Rewriting to eliminate redundant data preparation and predicate evaluation, reducing instruction construction and dispatch. It also designs Circuit Reduction to cancel inverse gate pairs, merge same-axis rotations, and combine successive single-qubit operations before backend execution. Both of them can efficiently reduce the execution cost.

\subsection{Version Management and State Reuse}
TimeArr stores historical arrays through backward deltas and accelerates version retrieval with skip links~\cite{timearr}, while Decibel shares data across relational branches~\cite{decibel}. TQSim partitions circuits and reuses intermediate subcircuit results across noisy simulations~\cite{tqsim}; qTask restricts incremental simulation to partitions affected by circuit modifications~\cite{qtask}. Heisenberg simulation propagates observables through circuits, with Pauli representations supporting expectation estimation~\cite{gottesman,rall}. \sys{} manages quantum state versions using shared checkpoints and operation histories. It reconstructs historical states through forward replay or inverse execution. The reconstructed states can serve as inputs to subsequent circuit execution. It also designs a cost-based quantum data reuse strategy for historical queries, which selects the better of inverse reconstruction and observable propagation on a retained successor to reduce the time cost, considering state-vector size, inverse gate operations, and Pauli-term processing.
\section{Conclusion}
\label{sec:conclusion}
This paper proposes \sys{}, a hybrid database supporting quantum data management. we proposes a unified object model and  QSQL interfaces for quantum data processing. By introducing hybrid optimization on operator combinations using dynamic programming and designing three-level instructions intermediate representation \sys{}-IR, \sys{} integrates quantum operations into database processing, enabling the database to support native quantum algorithms. Moreover, \sys{} designs the special quantum data version management strategy  and manage quantum data through shared checkpoints and operation histories. \sys{} also supports historical quantum state reconstruction by forward replay or inverse execution from retained successors. Extensive results on real datasets show that \sys{} achieves up to  $2.14\times$ speedup on relational queries and up to $25.86\times$ speedup in quantum data processing latency over the evaluated baselines.
\section*{AI-Generated Content Acknowledgment}
We use generative AI solely for language refinement, including grammar checks and language polishing. The authors are fully accountable for the content.

\bibliographystyle{IEEEtran}
\bibliography{references}

@inproceedings{timearr,
  author       = {Emad Soroush and Magdalena Balazinska},
  title        = {Time Travel in a Scientific Array Database},
  booktitle    = {Proc. IEEE Int. Conf. Data Eng. (ICDE)},
  year         = {2013},
  pages        = {98--109},
  doi          = {10.1109/icde.2013.6544817},
  url          = {https://doi.org/10.1109/icde.2013.6544817},
  publisher    = {IEEE}
}

@article{schoenberger2023join,
  author = {Manuel Sch{\"o}nberger and Stefanie Scherzinger and Wolfgang Mauerer},
  title = {Ready to Leap (by Co-Design)? {Join} Order Optimisation on Quantum Hardware},
  journal = {Proc. ACM Manag. Data},
  volume = {1},
  number = {1},
  pages = {92:1--92:27},
  year = {2023},
  doi = {10.1145/3588946}
}

@article{fankhauser2023mqo,
  author  = {Tobias Fankhauser and Marc E. Sol{\`e}r and
             Rudolf Marcel F{\"u}chslin and Kurt Stockinger},
  title   = {Multiple Query Optimization Using a Gate-Based Quantum Computer},
  journal = {IEEE Access},
  volume  = {11},
  pages   = {114031--114043},
  year    = {2023},
  doi     = {10.1109/ACCESS.2023.3324253}
}

@article{nayak2024split,
  author  = {Nitin Nayak and Tobias Winker and
             Umut {\c{C}}al{\i}ky{\i}lmaz and
             Sven Groppe and Jinghua Groppe},
  title   = {Quantum Join Ordering by Splitting the Search Space of {QUBO} Problems},
  journal = {Datenbank-Spektrum},
  volume  = {24},
  number  = {1},
  pages   = {21--32},
  year    = {2024},
  doi     = {10.1007/s13222-024-00468-3}
}

@inproceedings{liu2025bushy,
  author    = {Hanwen Liu and Abhishek Kumar and
               Federico Spedalieri and Ibrahim Sabek},
  title     = {Hybrid Quantum-Classical Optimization for Bushy Join Trees},
  booktitle = {Proc. 2nd Workshop Quantum Computing and Quantum-Inspired
               Technology for Data-Intensive Systems and Applications
               (Q-Data)},
  pages     = {20--24},
  year      = {2025},
  doi       = {10.1145/3736393.3736695}
}

@article{elnozahy2002rollback,
  author  = {E. N. Elnozahy and Lorenzo Alvisi and
             Yi-Min Wang and David B. Johnson},
  title   = {A Survey of Rollback-Recovery Protocols
             in Message-Passing Systems},
  journal = {ACM Comput. Surv.},
  volume  = {34},
  number  = {3},
  pages   = {375--408},
  year    = {2002},
  doi     = {10.1145/568522.568525}
}

@article{calikyilmaz2023opportunities,
  author  = {Umut {\c{C}}al{\i}ky{\i}lmaz and Sven Groppe and
             Jinghua Groppe and Tobias Winker and Stefan Prestel and
             Farida Shagieva and Daanish Arya and Florian Preis and
             Le Gruenwald},
  title   = {Opportunities for Quantum Acceleration of Databases:
             Optimization of Queries and Transaction Schedules},
  journal = {Proc. VLDB Endow.},
  volume  = {16},
  number  = {9},
  pages   = {2344--2353},
  year    = {2023},
  doi     = {10.14778/3598581.3598603}
}

@article{qiro,
  author = {David Ittah and Thomas H{\"a}ner and Vadym Kliuchnikov and Torsten Hoefler},
  title = {{QIRO}: A Static Single Assignment-based Quantum Program Representation for Optimization},
  journal = {ACM Transactions on Quantum Computing},
  volume = {3},
  number = {3},
  pages = {14:1--14:32},
  year = {2022},
  doi = {10.1145/3491247}
}

@inproceedings{quartz,
  author = {Mingkuan Xu and Zikun Li and Oded Padon and Sina Lin and Jessica Pointing and Auguste Hirth and Henry Ma and Jens Palsberg and Alex Aiken and Umut A. Acar and Zhihao Jia},
  title = {{Quartz}: Superoptimization of Quantum Circuits},
  booktitle = {Proc. ACM SIGPLAN Conference on Programming Language Design and Implementation (PLDI)},
  pages = {625--640},
  year = {2022},
  doi = {10.1145/3519939.3523433}
}

@article{netsquid,
  author       = {Tim Coopmans and Robert Knegjens and Axel Dahlberg and David Maier and Loek Nijsten and
                 Julio de Oliveira Filho and Martijn Papendrecht and Julian Rabbie and Filip
                 Rozp{\k{e}}dek and Matthew Skrzypczyk and Leon Wubben and Walter de Jong and Damian
                 Podareanu and Ariana Torres-Knoop and David Elkouss and Stephanie Wehner},
  title        = {{NetSquid}, a {NETwork Simulator for QUantum Information using Discrete events}},
  journal      = {Commun. Phys.},
  volume       = {4},
  year         = {2021},
  doi          = {10.1038/s42005-021-00647-8},
  note         = {{Art. no.} 164}
}

@article{q2o2025,
  author       = {Hanwen Liu and Federico Spedalieri and Ibrahim Sabek},
  title        = {A Demonstration of {Q$^2$O}: Quantum-Augmented Query Optimizer},
  journal      = {Proc. VLDB Endow.},
  volume       = {18},
  number       = {12},
  pages        = {5439--5443},
  year         = {2025},
  doi          = {10.14778/3750601.3750691},
  url          = {https://www.vldb.org/pvldb/vol18/p5439-liu.pdf}
}

@misc{nayak2026,
  author       = {Nitin Nayak and Tobias Winker and Umut {\c{C}}al{\i}ky{\i}lmaz and Jinghua Groppe and
                 Sven Groppe},
  title        = {Improved Join Order Optimization for Database Queries using Hybrid Quantum-Classical
                 Approaches for {QUBO} Problems},
  howpublished = {arXiv:2606.16247},
  year         = {2026},
  url          = {https://arxiv.org/abs/2606.16247}
}

@inproceedings{hopf2026,
  author       = {Patrick Hopf and Erick Ochoa Lopez and Yannick Stade and Damian Rovara and Nils
                 Quetschlich and Ioan Albert Florea and Josh Izaac and Robert Wille and Lukas
                 Burgholzer},
  title        = {Integrating Quantum Software Tools with(in) {MLIR}},
  booktitle    = {Proc. Supercomputing Asia and International Conference on High Performance Computing in
                 Asia Pacific Region (SCA/HPCAsia)},
  pages        = {42--54},
  year         = {2026},
  doi          = {10.1145/3773656.3773658},
  url          = {https://arxiv.org/abs/2601.02062}
}

@article{qdbo,
  author       = {Hanwen Liu and Abhishek Kumar and Federico Spedalieri and Ibrahim Sabek},
  title        = {{QDBO}: A Real-time Quantum-augmented Database System Optimizer},
  journal      = {Proc. VLDB Endow.},
  volume       = {19},
  number       = {11},
  pages        = {3606--3620},
  year         = {2026},
  doi          = {10.14778/3836663.3836712}
}

@inproceedings{qtask,
  author       = {Tsung-Wei Huang},
  title        = {{qTask}: Task-parallel Quantum Circuit Simulation with Incrementality},
  booktitle    = {Proc. IEEE Int. Parallel Distrib. Process. Symp. (IPDPS)},
  pages        = {746--756},
  year         = {2023},
  doi          = {10.1109/IPDPS54959.2023.00080}
}

@inproceedings{plank1995libckpt,
  author       = {James S. Plank and Micah Beck and Gerry Kingsley and Kai Li},
  title        = {{Libckpt}: Transparent Checkpointing under {Unix}},
  booktitle    = {Proc. USENIX Winter Tech. Conf.},
  address      = {New Orleans, LA, USA},
  year         = {1995},
  month        = jan,
  pages        = {213--223},
  url          = {https://web.eecs.utk.edu/~jplank/plank/papers/USENIX-95W.html}
}

@article{labaymora2024,
  author       = {Labay Mora, Adri{\`a} and {Ferreira da Silva}, Francisco and Stephanie Wehner},
  title        = {Reducing hardware requirements for entanglement distribution via joint
                 hardware-protocol optimization},
  journal      = {Quantum Sci. Technol.},
  volume       = {9},
  number       = {4},
  year         = {2024},
  doi          = {10.1088/2058-9565/ad57e9},
  note         = {{Art. no.} 045001}
}

@misc{entanglementdata2023,
  author = {{Ferreira da Silva}, Francisco},
  title        = {Replication data for: Reducing hardware requirements for entanglement distribution via
                 joint hardware-protocol optimization},
  howpublished = {4TU.ResearchData},
  year         = {2023},
  note         = {Dataset, version 1},
  doi          = {10.4121/0c6f100c-cf16-4fe1-81fe-afcafabdc7ca.v1},
  url          = {https://doi.org/10.4121/0c6f100c-cf16-4fe1-81fe-afcafabdc7ca.v1}
}

@inproceedings{burgholzer2026,
  author       = {Lukas Burgholzer and Daniel Haag and Yannick Stade and Damian Rovara and Patrick Hopf
                 and Robert Wille},
  title        = {The {MQT} Compiler Collection: A Blueprint for a Future-Proof Quantum-Classical
                 Compilation Framework},
  booktitle    = {Proc. Design, Automation \& Test in Europe Conf. (DATE)},
  year         = {2026},
  doi          = {10.23919/DATE69613.2026.11539504},
  url          = {https://arxiv.org/abs/2604.08674},
  pages        = {1--7}
}

@article{quantumcryptography,
  author       = {S. Pirandola and U. L. Andersen and L. Banchi and M. Berta and D. Bunandar and R.
                 Colbeck and D. Englund and T. Gehring and C. Lupo and C. Ottaviani and J. L. Pereira and
                 M. Razavi and J. S. Shaari and M. Tomamichel and V. C. Usenko and G. Vallone and P.
                 Villoresi and P. Wallden},
  title        = {Advances in Quantum Cryptography},
  journal      = {Adv. Opt. Photon.},
  volume       = {12},
  pages        = {1012--1236},
  year         = {2020},
  doi          = {10.1364/AOP.361502},
  number       = {4}
}

@article{trummer2016,
  author       = {Immanuel Trummer and Christoph Koch},
  title        = {Multiple Query Optimization on the {D-Wave 2X} Adiabatic Quantum Computer},
  journal      = {Proc. VLDB Endow.},
  volume       = {9},
  number       = {9},
  pages        = {648--659},
  year         = {2016},
  url          = {https://www.vldb.org/pvldb/vol9/p648-trummer.pdf},
  doi          = {10.14778/2947618.2947621}
}

@article{huang2022,
  author       = {Hsin-Yuan Huang and Michael Broughton and Jordan Cotler and Sitan Chen and Jerry Li and
                 Masoud Mohseni and Hartmut Neven and Ryan Babbush and Richard Kueng and John Preskill
                 and Jarrod R. McClean},
  title        = {Quantum Advantage in Learning from Experiments},
  journal      = {Science},
  volume       = {376},
  number       = {6598},
  pages        = {1182--1186},
  year         = {2022},
  doi          = {10.1126/science.abn7293}
}

@article{statsceb,
  author       = {Yuxing Han and Ziniu Wu and Peizhi Wu and Rong Zhu and Jingyi Yang and Liang Wei Tan
                 and Kai Zeng and Gao Cong and Yanzhao Qin and Andreas Pfadler and Zhengping Qian and
                 Jingren Zhou and Jiangneng Li and Bin Cui},
  title        = {Cardinality Estimation in {DBMS}: A Comprehensive Benchmark Evaluation},
  journal      = {Proc. VLDB Endow.},
  year         = {2022},
  volume       = {15},
  number       = {4},
  pages        = {752--765},
  doi          = {10.14778/3503585.3503586},
  url          = {https://www.vldb.org/pvldb/vol15/p752-zhu.pdf}
}

@article{luo2020yao,
  author       = {Xiu-Zhe Luo and Jin-Guo Liu and Pan Zhang and Lei Wang},
  title        = {{Yao.jl}: Extensible, Efficient Framework for Quantum Algorithm Design},
  journal      = {Quantum},
  volume       = {4},
  year         = {2020},
  doi          = {10.22331/q-2020-10-11-341},
  note         = {{Art. no.} 341}
}

@article{qulacs,
  author       = {Yasunari Suzuki and Yoshiaki Kawase and Yuya Masumura and Yuria Hiraga and Masahiro
                 Nakadai and Jiabao Chen and Ken M. Nakanishi and Kosuke Mitarai and Ryosuke Imai and
                 Shiro Tamiya and Takahiro Yamamoto and Tennin Yan and Toru Kawakubo and Yuya O. Nakagawa
                 and Yohei Ibe and Youyuan Zhang and Hirotsugu Yamashita and Hikaru Yoshimura and Akihiro
                 Hayashi and Keisuke Fujii},
  title        = {{Qulacs}: a fast and versatile quantum circuit simulator for research purpose},
  journal      = {Quantum},
  volume       = {5},
  year         = {2021},
  doi          = {10.22331/q-2021-10-06-559},
  note         = {{Art. no.} 559}
}

@article{lamb2024ising,
  author       = {Lamb, Christopher and Tang, Yicheng and Davis, Robert and Roy, Ananda},
  title        = {Ising meson spectroscopy on a noisy digital quantum simulator},
  journal      = {Nat. Commun.},
  volume       = {15},
  year         = {2024},
  doi          = {10.1038/s41467-024-50206-2},
  note         = {{Art. no.} 5901}
}

@article{hai2025,
  author       = {Rihan Hai and Shih-Han Hung and Tim Coopmans and Tim Littau and Floris Geerts},
  title        = {Quantum Data Management in the {NISQ} Era},
  journal      = {Proc. VLDB Endow.},
  volume       = {18},
  number       = {6},
  pages        = {1720--1729},
  year         = {2025},
  doi          = {10.14778/3725688.3725701}
}

@inproceedings{stade2025qir,
  author       = {Yannick Stade and Lukas Burgholzer and Robert Wille},
  title        = {Towards Supporting {QIR}: Steps for Adopting the Quantum Intermediate Representation},
  booktitle    = {Proceedings of the SC '25 Workshops of the International Conference for High
                 Performance Computing, Networking, Storage and Analysis},
  year         = {2025},
  pages        = {1907--1915},
  publisher    = {Association for Computing Machinery},
  doi          = {10.1145/3731599.3767546},
  url          = {https://doi.org/10.1145/3731599.3767546}
}

@misc{rieger,
  author       = {Carla Rieger and Michele Grossi and Gian Giacomo Guerreschi and Sofia Vallecorsa and
                 Martin Werner},
  title        = {Operational Framework for a Quantum Database},
  howpublished = {arXiv:2405.14947},
  year         = {2024},
  note         = {Version 2, Feb. 12, 2025},
  url          = {https://arxiv.org/abs/2405.14947v2}
}

@inproceedings{li2024,
  author       = {Tuodu Li and Gongsheng Yuan and Chang Yao and Meng Shi and Ziyue Wang and Ling Qian and
                 Jiaheng Lu},
  title        = {Quantum Storage Design for Tables in {RDBMS}},
  booktitle    = {Proc. 2nd Int. Workshop Quantum Data Science and Management (QDSM)},
  year         = {2024},
  url          = {https://vldb.org/workshops/2024/proceedings/QDSM/QDSM.4.pdf}
}

@article{inferq,
  author       = {Andrei Ilinescu and Aadi Patwardhan and Rihan Hai},
  title        = {{InferQ}: A Database-Oriented Benchmark for Quantum Circuits Simulation},
  journal      = {Proc. ACM Manag. Data},
  volume       = {4},
  number       = {4},
  pages        = {1--26},
  year         = {2026},
  month        = sep,
  note         = {{Art. no.} 278},
  doi          = {10.1145/3837116},
  url          = {https://doi.org/10.1145/3837116}
}

@article{mlae,
  author       = {Yohichi Suzuki and Shumpei Uno and Rudy Raymond and Tomoki Tanaka and Tamiya Onodera
                 and Naoki Yamamoto},
  title        = {Amplitude estimation without phase estimation},
  journal      = {Quantum Inf. Process.},
  volume       = {19},
  number       = {2},
  year         = {2020},
  doi          = {10.1007/s11128-019-2565-2},
  note         = {{Art. no.} 75}
}

@misc{gottesman,
  author       = {Daniel Gottesman},
  title        = {The {Heisenberg} Representation of Quantum Computers},
  howpublished = {arXiv:quant-ph/9807006},
  year         = {1998},
  url          = {https://arxiv.org/abs/quant-ph/9807006}
}

@article{rall,
  author       = {Patrick Rall and Daniel Liang and Jeremy Cook and William Kretschmer},
  title        = {Simulation of Qubit Quantum Circuits via {Pauli} Propagation},
  journal      = {Phys. Rev. A},
  volume       = {99},
  number       = {6},
  year         = {2019},
  doi          = {10.1103/PhysRevA.99.062337},
  note         = {{Art. no.} 062337}
}

@inproceedings{tqsim,
  author       = {Meng Wang and Swamit Tannu and Prashant J. Nair},
  title        = {Accelerating Simulation of Quantum Circuits under Noise via Computational Reuse},
  booktitle    = {Proc. Int. Symp. Comput. Archit. (ISCA)},
  year         = {2025},
  pages        = {1539--1553},
  doi          = {10.1145/3695053.3730992}
}

@article{decibel,
  author       = {Michael Maddox and David Goehring and Aaron J. Elmore and Samuel Madden and Aditya
                 Parameswaran and Amol Deshpande},
  title        = {{Decibel}: The Relational Dataset Branching System},
  journal      = {Proc. VLDB Endow.},
  volume       = {9},
  number       = {9},
  pages        = {624--635},
  year         = {2016},
  doi          = {10.14778/2947618.2947619}
}

@inproceedings{grover,
  author       = {Lov K. Grover},
  title        = {A Fast Quantum Mechanical Algorithm for Database Search},
  booktitle    = {Proc. ACM Symp. Theory of Computing (STOC)},
  pages        = {212--219},
  year         = {1996},
  doi          = {10.1145/237814.237866}
}

@misc{qaoa,
  author       = {Edward Farhi and Jeffrey Goldstone and Sam Gutmann},
  title        = {A Quantum Approximate Optimization Algorithm},
  howpublished = {arXiv:1411.4028},
  year         = {2014},
  url          = {https://arxiv.org/abs/1411.4028}
}

@article{projectq,
  author       = {Damian S. Steiger and Thomas H{\"a}ner and Matthias Troyer},
  title        = {{ProjectQ}: An Open Source Software Framework for Quantum Computing},
  journal      = {Quantum},
  volume       = {2},
  year         = {2018},
  doi          = {10.22331/q-2018-01-31-49},
  note         = {{Art. no.} 49}
}

@article{quantumcompilation,
  author       = {Thomas H{\"a}ner and Damian S. Steiger and Krysta Svore and Matthias Troyer},
  title        = {A Software Methodology for Compiling Quantum Programs},
  journal      = {Quantum Sci. Technol.},
  volume       = {3},
  number       = {2},
  year         = {2018},
  doi          = {10.1088/2058-9565/aaa5cc},
  note         = {{Art. no.} 020501}
}

@inproceedings{qymera,
  author       = {Tim Littau and Rihan Hai},
  title        = {{Qymera}: Simulating Quantum Circuits using {RDBMS}},
  booktitle    = {Companion of the 2025 Int. Conf. Management of Data},
  pages        = {179--182},
  year         = {2025},
  doi          = {10.1145/3722212.3725126}
}

@IEEEtranBSTCTL{IEEEcontrol,
  CTLuse_forced_etal = {yes},
  CTLmax_names_forced_etal = {6},
  CTLnames_show_etal = {1}
}

@misc{qute,
  author       = {Muzhi Chen and Xuanhe Zhou and Wei Zhou and Bangrui Xu and Surui Tang and Guoliang Li
                 and Bingsheng He and Yeye He and Yitong Song and Fan Wu},
  title        = {{Qute}: Towards Quantum-Native Database},
  year         = {2026},
  howpublished = {arXiv:2602.14699},
  url          = {https://arxiv.org/abs/2602.14699}
}

@inproceedings{dedalus,
  author       = {Emmanouil Limnaios and Markos Stergiopoulos and George T. Stamatiou and Efthymios
                 Papageorgiou and Vasilis Efthymiou and Dimitris Loupas and Dimitrios Tsourounis and
                 Kostas Blekos and Aggelos Tsikas and Dimitris Plexousakis and Kostas Magoutis and Yannis
                 Tzitzikas and Haridimos Kondylakis},
  title        = {{DEDALUS}: A Quantum-Enhanced End-to-End Framework for Cost-Aware Join Order
                 Optimization with Search Space Pruning},
  booktitle    = {Proc. Int. Conf. Extending Database Technology (EDBT)},
  year         = {2026},
  pages        = {716--719},
  doi          = {10.48786/edbt.2026.63}
}

@article{overeem2021eventsourcing,
  author       = {Michiel Overeem and Marten Spoor and Slinger Jansen and Sjaak Brinkkemper},
  title        = {An Empirical Characterization of Event Sourced Systems and Their Schema
                 Evolution---Lessons from Industry},
  journal      = {J. Syst. Softw.},
  volume       = {178},
  year         = {2021},
  note         = {{Art. no.} 110970},
  doi          = {10.1016/j.jss.2021.110970},
  url          = {https://arxiv.org/abs/2104.01146}
}
\end{document}